\documentclass[twocolumn,showkeys,preprintnumbers,amsmath]{revtex4}

\usepackage[usenames]{color}
\usepackage{graphicx}
\usepackage{epsfig}
\usepackage{dcolumn}
\usepackage{bm}
\usepackage{latexsym}
\usepackage{amsfonts}
\usepackage{amssymb}

\newcommand{\beq}{\begin{equation}}
\newcommand{\eeq}{\end{equation}}
\newcommand{\beqa}{\begin{eqnarray}}
\newcommand{\eeqa}{\end{eqnarray}}
\newcommand{\bea}{\begin{array}}
\newcommand{\ena}{\end{array}}

\begin{document}
\title{Black hole solutions coupled to Born-Infeld electrodynamics \\
with derivative corrections}
\author{Takashi Tamaki}
\email{tamaki@tap.scphys.kyoto-u.ac.jp}
\affiliation{Department of Physics, Kyoto University, 
606-8501, Japan}

\date{\today}
\begin{abstract}
We investigate black hole solutions in the Einstein-Born-Infeld system. 
We clarify the role played by derivative corrections to the Born-Infeld (BI) action. 
The qulitative differences from the case without derivative corrections are: (i) 
there is no particlelike solution. (ii) the existence of the inner horizon is 
restricted to the near extreme solutions. (iii) contribution of 
the BI parameter $b$ to the gravitational mass and the Hawking temperature works in 
the opposite direction. 
\end{abstract}
\keywords{blh,gra}

\preprint{KUNS-1878}

\maketitle

\section{Introduction}

Recently, much attention has been paid for Born-Infeld (BI) nonlinear 
electrodynamics \cite{Born} which naturally arises as a result of string 
corrections \cite{Frad}. One of the reason is that the world volume action of a D-brane 
is described by the BI action in the weak coupling limit \cite{brane}. Thus, the BI action 
has played important roles in D-brane physics. 

Moreover, since it is important to describe high energy region, particlelike solutions and black holes 
coupled to BI electrodynamics under the assumptions of static and spherically symmetric 
metric have been considered in the literature \cite{Demi,Oliveira,Fernando}. 
The existence of particlelike solutions shows the difference from the usual electrodynamics and these 
have been regarded as one of the realization of the electromagnetic geon in the 
literature \cite{Demi}. Thermodynamic properties and internal structure of these black holes are also 
changed from the Reissner-Nordstr\"om (RN) black holes. The black hole singularity in the 
BI action is weaked from that of RN black holes. 
These solutions were also extended to the case for non-Abelian BI field 
\cite{Gal,Dya,Grandi,Tripathy} or the case coupled to the dilaton \cite{TT,Clement,Ida}. 

Although above results seem to suggest that properties of charged objects differ from those 
in the usual electrodynamics, there is a problem one should consider. 
Unfortunately, since the BI action is the tree-level action derived by assuming 
the constancy of the field, we must consider derivative corrections 
if the field varies for the theoretical consistency \cite{Tsey,Andreev}. 
We ask whether or not above features are changed or maintained qualitatively 
if we include these derivative corrections. This is our purpose and we 
reveal some properties due to these corrections. 

\section{Model and Basic Equations}
We begin with the action 
\beqa
\hspace{-5mm}&&S  =  \displaystyle\int d^{4}x \sqrt{-g}\left[
\frac{R}{16\pi G}+\frac{b}{4\pi}\left(1-\sqrt{1+\frac{F_{ab}F^{ab}}{2b}}
\right)-\right.  \nonumber  \\
\hspace{-5mm}&&\frac{1}{192\pi^2 b^{3/2}}\left(F_{ab}F^{ab}\nabla_{a}F_{bc}\nabla^{a}F^{bc}+
8F_{kl}F^{lm}\nabla_{a}F_{mn}\nabla^{a}F^{nk}\right.
\nonumber  \\
\hspace{-5mm}&&\left.\left.
-4F_{la}F^{lb}\nabla_{b}F_{mn}\nabla^{a}F^{mn}
\right)\right],     \label{BI} 
\eeqa
where $G$ is the gravitational constant. The BI parameter $b$ can be written in terms of 
inverse string tension $\alpha'$ as $b=(2\pi\alpha ' )^{-2}$. We neglected the 
higher order derivative terms or the dilaton, for simplicity. Of course, although these should be 
included in the theoretical view point, they would interupt to interpret the role of the derivative 
corrections analytically. For this reason, we consider simplified model at present. 
Notice that the action (\ref{BI}) reduces to the Einstein-Maxwell system 
in the limit $b\to\infty$. 

We assume that a space-time is static and spherically 
symmetric, in which the metric is written as 
\beqa
ds^{2}=-f(r)e^{-2\delta (r)}dt^{2}+
f(r)^{-1}dr^{2}+r^{2}d\Omega^{2}, 
\label{metric}
\eeqa
where $f(r):=1-2Gm(r)/r$. We consider the magnetically charged case 
$F_{\theta\phi}=-Q_{m}\sin \theta$. 

Under the above assumptions, the basic equations are
\beqa
\bar{m}'&=&\bar{b}\lambda_{H}^{2}\left(\sqrt{\bar{r}^{4}+
\frac{Q_{m}^{2}}{\bar{b}\lambda_{H}^{4}}}-\bar{r}^{2}\right)
+fD\ ,  
\label{m} \\
\delta '&=&-\frac{2D}{\bar{r}}\ , 
\label{del} 
\eeqa
where $'=d/d\bar{r}$ and
\beqa
D&=&\frac{5Q_{m}^{4}}{3\pi\bar{b}^{3/2}\bar{r}^{8}\lambda_{H}^{8}}\ .
\label{derivative}
\eeqa
We have introduced the following dimensionless variables:
\beqa
\bar{b}:&=&G^{2}b,\ \ \bar{r}:=r/r_{H},\\ 
\bar{m}:&=&Gm/r_{H},\ \ \lambda_{H}=r_{H}/\sqrt{G}, 
\eeqa
where $r_{H}$ is the horizon radius. 

The term originated from the derivative terms only appears in the term 
$D$ in Eq.~(\ref{derivative}). Notice that particlelike solutions does not exist 
in which case $m(0)=0$ is assumed for the regularity at the origin. 
However, $m'(0)$ diverges except the case $f(0)=0$ which is not generic because of 
$\bar{m}'(0)=\sqrt(\bar{b})Q_{m}=1/2$. Even if $f(0)=0$ is satisfied, it is not enough 
because of $D\propto \bar{r}^{-8}$. Thus, one of the basic properties is 
altered due to the term $D$. 

Below, we consider black hole solutions. 
We assume the regular event horizon (EH) at $r=r_{H}$. 
\beqa
Gm_{H}&=&\frac{r_{H}}{2},\;\; \delta_H< \infty .
\label{mth}
\eeqa
The variables with subscript $H$ are evaluated 
at the horizon. We also assume the boundary conditions at spatial infinity as
\beqa
\sqrt{G}m(\infty)=: M=const.,\ \ \delta (\infty)=0, 
\label{atinf}
\eeqa
which means that the space-time is asymptotically flat. Here, we 
chose $Q_{m}=0.1$ for simplicity. 

We define the inner horizon (IH) as $r_{\rm in}$ which satisfies 
\beqa
f(r_{\rm in})=0\ (r_{\rm in}<r_{H}). 
\label{inner}
\eeqa
Since we can find out that $\delta$ has finite value if the integration does not 
include the origin from Eq.~(\ref{del}), this is verified. 
We write as $\lambda_{\rm in}:=r_{\rm in}/\sqrt{G}$.  

\section{Properties of BI black holes} 
First, we summarize the properties of black hole with no derivative terms, which we 
denote BI black holes. Solutions can be expressed as $\delta =0$ and \cite{Demi,Oliveira}
\beqa
&&m(r)=m_0+\frac{bQ_m^{2}}{3}\left[
\frac{r}{r^2+\sqrt{r^4+bQ_m^2}} \right.  
\nonumber \\
&& \left. +\frac{1}{\sqrt{b}Q_m}F\left(\frac{1}{\sqrt{2}},
\arccos \frac{\sqrt{b}Q_m -r^2}{\sqrt{b}Q_m +r^2}
\right)\right]   , 
\label{eliptic}
\eeqa
where $F(k,\varphi)$ is the elliptic function of the first kind. 
The constant $m_0$ is the mass inside EH. 

We exhibit the relations between the horizon radius $\lambda_{H}$ and the 
gravitational mass $M$ in Fig.~\ref{default} (a). BI and RN black holes are plotted in 
dotted lines and a solid line, respectively. 
To understand this diagram, we comment on the extreme solution where 
$\bar{m}_{H}'=1/2$ is satisfied. Then, we obtain 
\beqa
\bar{b}\lambda_{H}^{2}=\bar{b}Q_{m}^{2}-\frac{1}{4}\ ,
\label{extreme}
\eeqa
from Eq.~(\ref{m}). This is not satisfied for $\sqrt{\bar{b}}Q_{m} \leq 1/2$, 
i.e., $\bar{b}\leq 25$ for $Q_{m}=0.1$. Thus, there is no extreme solution  
in this case. For this reason, solutions exist until the limit $\lambda_{H}\to 0$. 
For the solutions with $\sqrt{\bar{b}} >25$, lower bound of $\lambda_{H}$ is 
determined by the extreme condition (\ref{extreme}). 

We notice that if we fix $\lambda_H$, the mass of the BI black holes 
monotonically increases with $\bar{b}$. We can confirm this by differentiating 
Eq.~(\ref{m}) by $\bar{b}$, i.e., 
\newtheorem{remark}{Remark}
\begin{remark}
If we fix $\lambda_H$ and $\bar{r}$, $\bar{m}'$ increases as $\bar{b}$. 
\end{remark}

We also exhibit the relations between the inverse Hawking temperature $1/T$ and the 
gravitational mass $M$ in Fig.~\ref{default} (b). It is convenient to write down the 
temperature as~\cite{Visser}
\beqa
T=\frac{e^{-\delta_H}}{4\pi \lambda_H}(1-2\bar{m}_{H}')\ .
\label{temperature}
\eeqa
If the weak energy condition is satisfied, both $\delta_{H}$ and $\bar{m}_{H}'$ 
are positive. This means that black holes including matter fields have lower 
temperature than the Schwarzschild black hole~\cite{Visser}. 

For the BI case, since $\delta_H =0$, 
the behavior of $T$ depends only on $\lambda_{H}$ and $\bar{m}_{H}'$. 
Then by {\bf Remark 1}, the temperature decreases as $\bar{b}$ increases 
for fixed $\lambda_{H}$. This is also reflected for fixed $M$, since $M$ increases 
with $\lambda_{H}$ as shown in Fig.~\ref{default} (a). 
Then, the lines shifts to the upperward as $\bar{b}$ increases as shown in Fig.~\ref{default} (b). 
Notice that solutions exist until $\lambda_{H}\to 0$ for $\bar{b}\leq 25$. 
Thus, $T$ diverges for $\bar{b}=20$ in this limit while it does not for $\bar{b}=25$ since 
$\bar{m}_{H}'=1/2$ is satisfied. 

\section{Comparison of BI and BID black holes} 
\begin{figure}[htbp]
\psfig{file=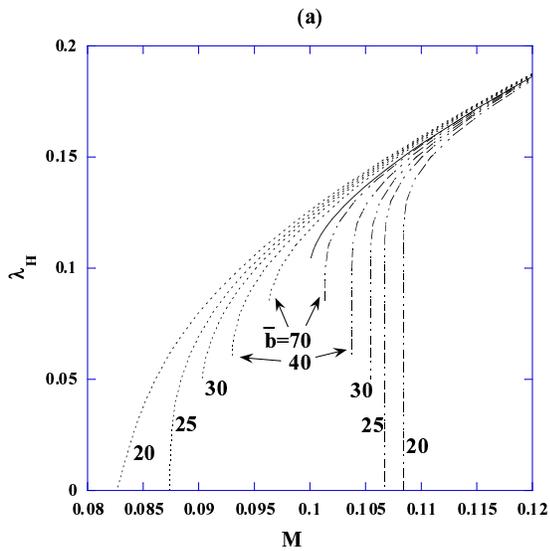,width=3.5in}
\psfig{file=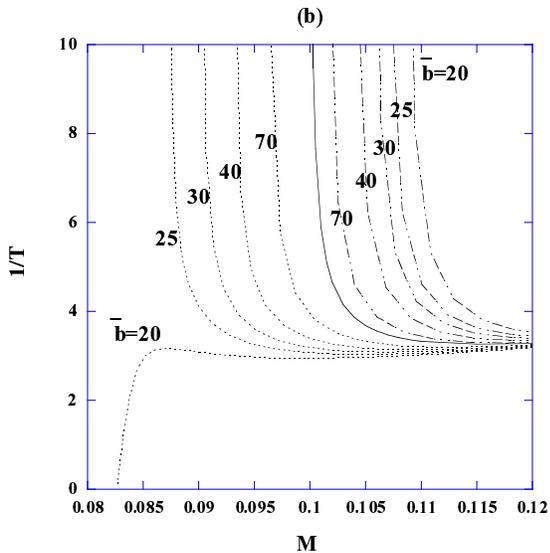,width=3.5in}
\caption{(a) $M$-$\lambda_{H}$ (b) $M$-$1/T$ for BID, BI and RN black holes are 
plotted in dot-dashed lines, dotted lines, and a solid line, respectively. 
\label{default} }
\end{figure}

Next, we compare BI black holes with black holes including derivative terms, 
which we denote BID black holes. Properties of BID black holes are shown in dot-dashed 
lines in Figs.~\ref{default}.  In the limit $\bar{b}\to\infty$, both BI and 
BID black holes converge to the RN black holes. In Fig.~\ref{default} (a), we find that 
the lower bound of $\lambda_{H}$ coincides for fixed $\bar{b}$ in both cases. 
This is due to the fact that the extreme condition (\ref{extreme}) coincides in both cases 
since $f=0$ at the horizon. 

However, the mass of BID black holes increases by reducing $\bar{b}$. 
This is a consequence of the term $D$ in Eq.~(\ref{m}) which is proportional 
to $\bar{b}^{-3/2}$. This is one of qualitative differences from the BI case. 
Let us also consider the behavior in Fig.~\ref{default} (b). We find that $\bar{b}$ 
works in the opposite direction in these cases as in Fig.~\ref{default} (a). 
Because of $f_{H}=0$, the term $D$ in Eq.~(\ref{m}) is not relevant in this case. 
The crucial factor is $\delta_{H}\propto\bar{b}^{-3/2}$. 
As a result, the temperature decreases as $\bar{b}$ decreases. 
Thus, the evaporation process of charged black holes would be quite different from 
the BI case even if the value of $\bar{b}$ is fixed.  

\begin{figure}[htbp]
\psfig{file=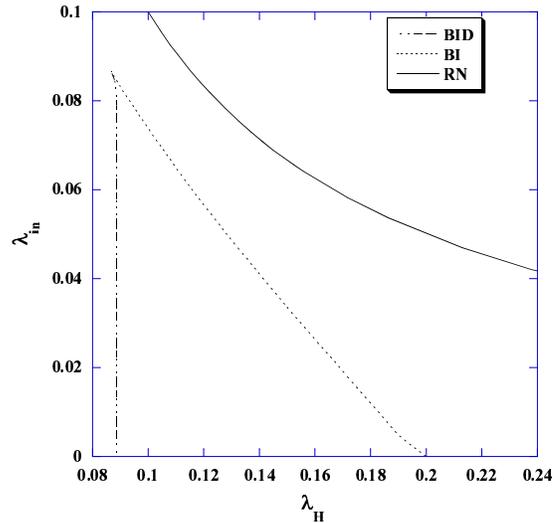,width=3.5in}
\caption{$\lambda_H$-$\lambda_{\rm in}$ for solutions with $\bar{b}=100$ 
and the RN solution. 
\label{rin} }
\end{figure}

We turn our attention to the inner structure of black holes. We show the relation 
between $\lambda_H$ and the IH 
$\lambda_{\rm in}$ for BI and BID black holes with $\bar{b}=100$ and RN solution 
in Fig.~\ref{rin}. Qualitative difference in these three types of solutions appears. 

\begin{figure}[htbp]
\psfig{file=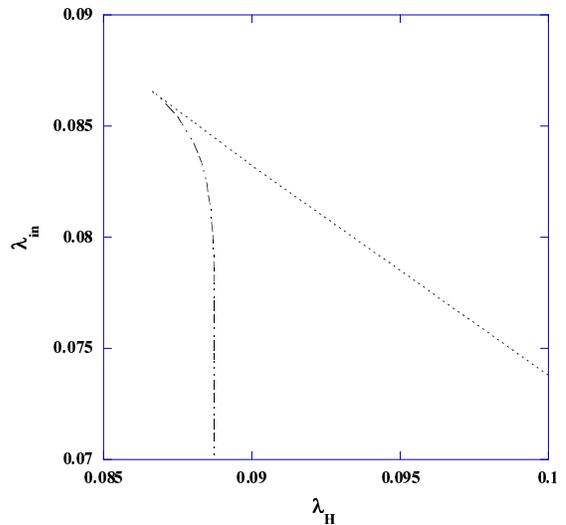,width=3.5in}
\caption{Magnification of Fig.~\ref{rin}.
\label{rin2} }
\end{figure}

First, we compare RN and BI case for fixed $\lambda_{H}$. We notice that $\lambda_{\rm in}$ 
for BI black holes is smaller than that for RN black holes. We can understand this behavior 
as $\bar{b}$ reduces the effect of charge. Because of {\bf Remark 1}, BI black hole approaches 
Schwarzschild black hole as $\bar{b}$ decreases. 

For the BID case, we may think that it is strange, since almost vertical line appears 
in Fig.~\ref{rin}. It is not a numerical artifact. To confirm it, 
we also exhibit a magnification of Fig.~\ref{rin} 
in Fig.~\ref{rin2}. As we stated above, lower bound of $\lambda_H$ coincides in these cases. 
For this reason, lines merge in the extreme limit. 
As the solutions deviate from the extreme limit, difference becomes outstanding. 

\begin{figure}[htbp]
\psfig{file=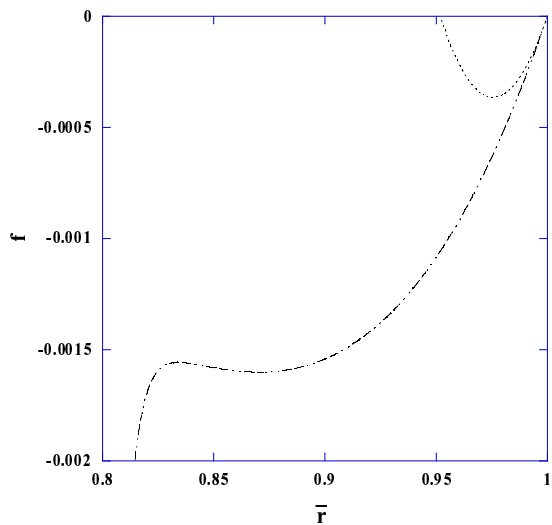,width=3.5in}
\caption{The metric function $f$ inside the horizon for the solutions with $\bar{b}=100$ 
and $\lambda_{H}=0.088744795$. BI and BID black holes are shown by 
dotted line and dot-dashed line, respectively. \label{r-grr} }
\end{figure}

To understand this, we should see that the 
term $D$ in Eq.~(\ref{m}) is proportional to $\bar{r}^{-8}$ and is multiplied by 
the factor $f$. $f$ in front of $D$ disturbs the contribution of $D$ at the vicinity 
of the EH. If $f$ remains small enough as the near extreme solution, the BID case is 
close to the BI case. However, this is very sensitive to the value $f$. 
Let us see this feature by viewing  Fig.~\ref{r-grr}. 
Difference from the BI case is small near the EH. However, if $|f|$ 
becomes large as we proceed toward inside, the result becomes quite different from the 
BI case. Since this is also sensitive to 
$\lambda_H$, small deviation of $\lambda_H$ greatly affects $\lambda_{\rm in}$. 
Thus, the almost vertical line in Fig.~\ref{rin2} appears. 
This tendency becomes more 
clear for larger $\bar{b}$, since $D\propto \bar{b}^{-3/2}$. 

We can also show that there is only one IH at most. We show contradiction by 
assuming that there are two IH. For this purpose, we write down $f'$ as 
\beqa
f'=\frac{2(\bar{m}-\bar{m}'\bar{r})}{\bar{r}^2}\ . 
\label{judge}
\eeqa
At the first IH $\lambda_{\rm in}$, we 
notice that $f=0$ (i.e., $\bar{r}=2\bar{m}$) and $f'<0$ which mean $\bar{m}'(\lambda_{\rm in} )>1/2$ 
from Eq.~(\ref{judge}). While, we should have $f'>0$ at the second inner horizon 
$\lambda_{\rm in2}$ which means $\bar{m}'(\lambda_{\rm in2} )<1/2$. 
This means $\bar{m}'(\lambda_{\rm in} )>\bar{m}'(\lambda_{\rm in2} )$ for 
$\lambda_{\rm in}>\lambda_{\rm in2}$. 
However, it is impossible if we notice that the first term of r.h.s. in Eq.~(\ref{m}) 
monotonically decreases with $\bar{r}$. 
(Notice that the second term of r.h.s. in Eq.~(\ref{m}) is not relevant to this proof 
because of $f=0$.) 
Thus, they have only one IH at most.

\section{Conclusion and discussion}

We have investigated the effect of derivative correction terms in BI action to 
black holes and found that IH exists only near the extreme solution 
in contrast with BI black holes. 
We also found that the BI parameter works in the opposite direction from that for 
the BI black holes.  

We comment on the stability of our solutions. In our previous papers, 
we considered a stability criterion using catastrophe theory \cite{cata}. 
From the result, stability change occurs at $d(1/T)/dM\to\infty$ 
\cite{Katz,torii,BD,tamaki}. In our result, we can find that there is no 
such point. Thus, BI and BID solutions would be stable. 

As a future work, higher order derivative correction should be included. 
If we surmize the result from this paper, above tendency would be strengthened. 
As we investigated before \cite{TT}, other fields such as a dilaton field or an axion 
field might also be important. If we consider the coupling of these fields to the 
derivative term, they may change stability and want to investigate in future. 

\section*{Ackowledgement}
Special Thanks to J. Soda for continuous encouragement. 
This work was supported in part by Grant-in-Aid for Scientific Research Fund of the
Ministry of Education, Science, Culture and Technology of Japan, 2003, No.\ 154568. 
This work was also supported in part by a Grant-in-Aid for the 21st Century 
COE ``Center for Diversity and Universality in Physics".



\end{document}